\documentclass[11pt]{article}

\usepackage{fullpage,amsmath,amsthm}
\usepackage{amssymb,latexsym}

\usepackage{times}

\newcommand{\myomit}[1]{}

\newcommand{\ket}[1]{|#1\rangle}

\newcommand{\ketbra}[2]{|#1\rangle\langle#2|}
\newcommand{\inp}[2]{\langle{#1}|{#2}\rangle} 

\newcommand{\Tr}{\mbox{\rm Tr}}

\newcommand{\BIJ}{\textsc{Bijection Test}}
\newcommand{\MA}{\textsc{Matching Test}}
\newcommand{\QMS}{\textsc{QMIP$^*$}}
\newcommand{\QMIP}{\textsc{QMIP}}
\newcommand{\GM}{\textsc{gap-3DM}}
\newcommand{\gmlong}{\textsc{gap-3D-MATCHING}}
\newcommand{\Pe}{\textsc{P}}
\newcommand{\NP}{\textsc{NP}}
\newcommand{\QIP}{\textsc{QIP}}
\newcommand{\IP}{\textsc{IP}}
\newcommand{\EXP}{\textsc{EXP}}
\newcommand{\NEXP}{\textsc{NEXP}}
\newcommand{\MIP}{\textsc{MIP}}

\newcommand{\PCP}{\textsc{PCP}}
\newcommand{\N}{\ensuremath{\mathbb{N}}}
\newtheorem{theorem}{Theorem}

\newtheorem{lemma}[theorem]{Lemma}
\newtheorem{conjecture}[theorem]{Conjecture}
\newtheorem{claim}[theorem]{Claim}
\newtheorem{fact}[theorem]{Fact}
\newtheorem{corollary}[theorem]{Corollary}
\newtheorem{definition}[theorem]{Definition}

\newcommand{\be}{\begin{eqnarray}}
\newcommand{\ee}{\end{eqnarray}}

\newcommand{\Sum}[2]{\ensuremath{\textstyle{\sum\limits_{#1}^{#2}}}}

\newcommand{\eps}{\varepsilon}
\newcommand{\ph}{\ensuremath{\varphi}}

\begin{document}

\title{On the Power of Entangled Quantum Provers}
\author{
Julia Kempe\thanks{Supported in part by ACI S\'ecurit\'e Informatique SI/03 511 and ANR AlgoQP grants of the
French Research Ministry, and also partially supported by the European Commission under the Integrated
Project Qubit Applications (QAP) funded by the IST directorate as Contract Number 015848.}\\CNRS \&\ LRI\\
Univ.~de Paris-Sud, Orsay
 \and
Thomas Vidick\thanks{Work done while at LRI, Univ. de Paris-Sud, Orsay.}\\
DI, \'Ecole Normale Sup\'erieure\\
Paris
 }
\date{}

 \maketitle

\begin{abstract}
We show that the value of a general two-prover quantum game cannot be computed by a semi-definite program of
polynomial size (unless \Pe=\NP), a method that has been successful in more restricted quantum games. More
precisely, we show that proof of membership in the NP-complete problem $\gmlong$ can be obtained by a
$2$-prover, $1$-round quantum interactive proof system where the provers share entanglement, with perfect
completeness and soundness $s=1-2^{-O(n)}$, and such that the space of the verifier and the size of the
messages are $O(\log n)$. This implies that $\QMS_{\log n,1,1-2^{-O(n)}} \nsubseteq \Pe$ unless $\Pe = \NP$
and provides the first non-trivial lower bound on the power of entangled quantum provers,
%
albeit
with an exponentially small gap. The gap achievable by our proof system might in fact be larger, provided a
certain conjecture on almost commuting versus nearly commuting projector matrices is true.
\end{abstract}

\section{Introduction}

Multi-prover interactive proof systems have played a tremendous role in classical computer science, in
particular in connection with probabilistically checkable proofs (PCPs). The discovery of the considerable
expressive power of two-prover interactive proof systems, as expressed by the relation  $\MIP = \NEXP$
\cite{BFL91}, prompted a systematic study of the precise amount of resources (the randomness used by the
verifier, and the amount of communication between him and the provers) necessary to maintain this
expressivity. These investigations culminated in a new characterization of \NP, $\NP=\PCP(O(\log n),O(1))$
\cite{ALMSS,AS92}, known as the \emph{PCP Theorem}. This characterization has had wide-ranging applications,
most notably in the field of hardness of approximation, where it is the basis of almost all known results.

\myomit{Much of the study of these systems occurred before the mid-nineties, when quantum information was
less known in the theoretical computer science community and quantum strategies were not considered.}

The study of quantum interactive proofs was initiated by Watrous, who was the first to systematically study
proof systems with {\em one} prover, whose power is only limited by the laws of quantum mechanics and who
communicates quantum messages with a polynomially bounded quantum verifier (the class \QIP). Kitaev and
Watrous showed \cite{KitWat00} that $\QIP(3)$, the class of quantum interactive proofs with $3$ rounds can
simulate all of \QIP\ and is contained in the class \EXP, i.e. $\IP \subseteq \QIP=\QIP (3) \subseteq \EXP$.
The proof of the last inclusion uses the fact that the maximization task of the prover can be written as a
semi-definite program (SDP) of exponential size together with the fact that there are efficient algorithms to
compute their optimum \cite{VandenbergheBoyd:sdp,GLS:sdp}. Moreover, Raz \cite{Raz:pcp} showed that the PCP
theorem combined with quantum information can have surprising results in complexity theory. It would be
interesting to formulate a purely quantum PCP theorem, which could arise from the in-depth study of quantum
\emph{multi-prover} interactive proof systems.

When considering interactive proof systems with multiple provers, the laws of quantum mechanics enable us to
introduce an interesting new twist, namely, we can allow the provers to share an arbitrary (a priori)
entangled state, on which they may perform any local measurements they like to help them answer the
verifier's questions. This leads to the definition of the classes  $\MIP^*$ (communication is classical and
provers share entanglement), \QMIP\ (communication is quantum, but provers do not share entanglement) and
\QMS\ (communication is quantum and provers share entanglement). Kobayashi and Matsumoto \cite{KoMa03} showed
that $\QMIP = \MIP$, but the question of how entanglement influences the power of such proof systems remains
wide open.\footnote{It is still true that $\QMIP^*\subseteq \MIP$ when the provers share only a polynomial
amount of entanglement.} The fact that entanglement can cause non-classical correlations is a familiar idea
in quantum physics, introduced in a seminal 1964 paper by Bell \cite{Bell}. It is thus a natural question to
ask what the expressive power of entangled provers is.

The only recent result in this direction is by Cleve et al. \cite{CleveHTW04}, who show, surprisingly, that
$\oplus \MIP^*(2,1) \subseteq \EXP$, where $\oplus \MIP^*(2,1)$ is the class of one-round classical
interactive proofs where the two provers are allowed to share some arbitrary entangled state, but reply only
a bit each, and the verifier bases his decision solely on the XOR of the two answer bits.\footnote{This
result was recently strengthened by Wehner \cite{Wehner:MIP}, who showed that $\oplus \MIP^*(2,1) \subseteq
\QIP(2)$.} This should be contrasted with the corresponding classical class without entanglement: it is known
that $\oplus \MIP(2,1)=\NEXP$ due to work by H\aa stad \cite{Has01}. The inclusion $\oplus \MIP^*(2,1)
\subseteq \EXP$ follows from the fact that the maximization problem of the two provers can be written as an
SDP. More precisely, there is an SDP relaxation with the property that its solutions can be translated back
into a protocol of the provers. This is possible using an inner-product preserving embedding of vectors into
two-outcome observables due to Tsirelson \cite{tsirelson}.

It is a wide open question whether it is true that $\MIP^* \subseteq \EXP$ or even $\QMS \subseteq \EXP$. Is
it possible to generalize Tsirelson's embedding to study proof systems where the answers are not just one
bit? The semi-definite programming approach has proved successful in the only known characterizations of
quantum interactive proof systems: both for \QIP\ and for $\oplus \MIP^* (2,1)$ it was shown that the success
probability is the solution of a semi-definite program. Does this remain true when the provers reply more
than one bit, or when messages are quantum? There are  SDP relaxations for the success probability both in
the case of $\MIP^*$ and \QMS; is it possible that they are {\em tight}, implying inclusion in \EXP? Or could
it be on the contrary that $\NEXP \subseteq \QMS$?

In this paper we provide a  step towards answering these questions. We rule out the possibility that the
success probability of $\QMS$ systems can be given as the solution of a semi-definite program (unless $\Pe =
\NP$). Mainly for convenience, we state our results in the scaled down realm of polynomial time and
logarithmic communication. Here the analogous question is whether $\NP \subseteq \QMS_{\log n}$, where the
subscript $\log n$ indicates the corresponding proof system with communication  and verifier's space
logarithmic in the input size $n$. Our main result is the following:
\begin{theorem}\label{thm:main}
$\NP \subseteq \QMS_{\log n,1,s}(2,1)$ with soundness $s=1-C^{-n}$ for some constant $C>1$. The verifier,
when given oracle access to the input, requires only space and time $O(\log n)$.
\end{theorem}
To our knowledge this is the first lower bound on the power of entangled provers. Note that even an
exponentially small gap between completeness and soundness is not at all a triviality in our setting. For
instance, it is not possible for the verifier to guess one of the exponentially many solutions, since he only
has a logarithmic amount of space and randomness. We believe that our result is significant for the following
reasons. First, we introduce novel techniques that exploit quantum messages and quantum tests {\em directly}.
Our approach is to give a $2$-prover, $1$-round protocol for an \NP-complete problem, \gmlong\ (\GM), where
the verifier sends quantum messages of length $\log n$ to each of the provers, who reply with messages of the
same length. This protocol truly exploits the fact that the messages are {\em quantum}, and does not seem to
work for classical messages. To give a vague intuition as to why quantum messages help, imagine that the
verifier wants to send a question $u$ from a set $U$ to the provers and to enforce that their answers $v$ are
given according to a bijection $v=\pi(u)$. He could exploit quantum messages by preparing the state
$\ket{\phi}=\sum_{u \in U} \ket{u}_A \ket{u}_B$ and sending one register to each of the provers. If the
provers are honest, the resulting state is $\sum_{u \in U} \ket{\pi(u)}_A \ket{\pi(u)}_B$; but of course,
since the original state is invariant under a bijection, this is equal to the state $\ket{\phi}$. Hence, even
not knowing $\pi$ the verifier can measure the received state in a basis containing $\ket{\phi}$ to get an
indication whether the provers are honest. We use variations of this idea, together with the SWAP test, to
derive conditions on the provers' behavior, forcing them to apply approximate bijections.

Second, we pinpoint the bottleneck for decreasing soundness, which is related to the question:
 \begin{quote}
{\em  Given $n$ pairwise almost commuting projectors, how well can we approximate them by $n$ commuting
projectors?}
 \end{quote}
More precisely we link the soundness to the scaling of $\delta$ in the following conjecture:
\begin{conjecture}\label{conj}
Let $P_1,\ldots,P_{m}$ be projectors and $D$ some diagonal matrix such that $\|D\|_F=1$ (where $\|\cdot \|_F$
is the Frobenius norm) and $\|(P_iP_j-P_jP_i)D\|_F^2 \leq \eps$ for all $i,j\in\{1,\ldots,m\}$. Then there
exist a $\delta \geq 0$, diagonal projectors $Q_1,\ldots , Q_m$, and a unitary matrix $U$, such that $\forall
i$ $\|(P_i-UQ_iU^\dagger)D\|_F^2 \leq \delta$.
\end{conjecture}
Along with Theorem \ref{thm:main} we show the following
\begin{corollary}\label{cor:main}
There are constants $C,C',C''>0$ such that if Conjecture \ref{conj} is true for $m=Cn$ and
$\delta=\delta(n,\eps)$ then $\NP \subseteq \QMS_{\log n,1,1-\eps'}$ for $\eps'$ such that
$\delta(n,C''\eps')\leq C'$.
\end{corollary}
In particular if $\delta=poly(n)\cdot \eps$ we get soundness $s=1-poly(n)^{-1}$ and if $\delta = \delta
(\eps)$ is constant (independent of $n$) we get constant soundness $s$, and in a scaled up version $\NEXP
\subseteq \QMS_{1,s}$ for constant $s$.\footnote{Note that proving Conjecture \ref{conj} for $D$ proportional
to the identity matrix would give the corresponding result for provers that share a maximally entangled
state.} We show in Lemma \ref{lem:diagonal} that Conjecture \ref{conj} is true for $\delta=2^{O(n)}\cdot
\eps$, which gives soundness $s=1-2^{-O(n)}$. We conjecture that Conjecture \ref{conj} is true for $\delta=n
\eps$.

Finally, our result has an important consequence: it shows that standard SDP techniques will not work to
prove that $\QMS \subseteq \EXP$ and that the success probability of quantum games cannot be computed by an
SDP that is polynomial in the size of the verifier and of the messages (unless \Pe=\NP). In the case of
$\QMS_{\log n}$ with a $\log n$-space verifier the SDP would have size polynomial in $n$.\footnote{Note that
the SDP depends on the instance $x$ of \GM, but can be constructed from $x$ in polynomial time.} It is well
known that there are polynomial time algorithms to find the optimum of such SDP's up to exponential
precision; in particular these algorithms could {\em distinguish} between success probability $1$ and
$1-2^{-O(n)}$ and hence they could solve \NP - complete problems.
\begin{corollary}
Quantum games with entangled quantum provers cannot be computed by an SDP that is polynomial in the dimension
of the messages and of the verifier.
\end{corollary}

Another related consequence of our result is that there is no generic way to prove $\QMS \subseteq \QIP$,
because our results imply $\QMS_{\log n,1,1-2^{-O(n)}} \nsubseteq \QIP_{\log n,1,1-2^{-O(n)}}$, where
$\QIP_{\log n}$ is the class of quantum interactive proofs with  communication and verifier's size of order
$\log n$. This is true for the same reason as before: there is a polynomial size SDP for the success
probability of $\QIP_{\log n}$ protocols.

\medskip

{\em Related work:} Ben Toner \cite{toner:personal} communicated to us existing attempts to show $\text{NP}
\subseteq \MIP^*_{\log n}$, which focus on showing that in the
case that there are a large number of provers, imposing classical correlations on their answers
can help restrain the nonlocal correlations that they exhibit to the point where they cannot cheat
 more than two classical unentangled provers. It is possible
by symmetrization  to obtain  a relation which has some resemblance to Conjecture \ref{conj} (although in the
operator norm, where the conjecture is false), where $\eps$ is inverse proportional to the number of provers
in the protocol. After the completion of this work, we have heard of related work showing that $\NP \subseteq
\MIP^*_{\log n,c,s}(3,1)$, independently by Ben Toner, and Hirotada Kobayashi and Keiji Matsumoto. We can
therefore also conclude that semidefinite programs cannot compute the value of games with three entangled
provers and classical communication. Furthermore we have just learned from Hirotada Kobayashi and Keiji
Matsumoto about another lower bound on two-prover quantum systems that shows $\textsc{IP}=\textsc{PSPACE }
\subseteq \QMS$ with inverse polynomial soundness; and the authors communicated to us that they were
currently working on possibly extending this to a statement on \NEXP\ with simply exponential gap.

\medskip

The structure of this paper is as follows: In Section \ref{sec:not} we introduce the necessary definitions
and notations and give the version of \GM\ we use. In Section \ref{sec:zero} we show that  \GM\ can be put
into a {\em zero-error} version of $\QMS_{\log n}(2,1)$. We then show in Section \ref{sec:sound} that the
{\em zero-error} requirement can be relaxed to soundness $1-2^{-O(n)}$ proving Theorem \ref{thm:main} and
Corollary \ref{cor:main}. In Section \ref{sec:rest} we elaborate on Conjecture \ref{conj} and briefly discuss
scaling-up to proving $\NEXP \subset \QMS_{1,s} (2,1)$.

\section{Preliminaries}\label{sec:not}

We assume basic knowledge of quantum computation~\cite{nielsen&chuang:qc} and of classical interactive proof
systems \cite{lund:ip}. The relevant classes of quantum interactive proof systems are defined as follows.

\begin{definition}
A $(n,r,m)$ classical (resp. quantum) interactive proof system is given by a polynomial-time classical (resp.
quantum) circuit (the verifier V) that runs in space $O(m)$. V interacts with $n$ infinitely powerful quantum
provers through $n$ special classical (resp. quantum) channels. The verifier is allowed to communicate at
most $O(m)$ bits (resp. qubits) in a maximum of $r$ rounds of interaction through his communication channels.

Let $\MIP^*_{m,c,s}(n,r)$ (resp. $\QMS_{m,c,s}(n,r)$) denote the class of languages $L$ such that there
exists a $(n,r,m)$ classical (resp. quantum) interactive proof system  such that
\begin{itemize}
\item $\forall x\in L$, there exist $n$  provers who share a $n$-partite state $\ket{\Psi}$ such that the
interaction between V and the provers results in the verifier accepting with probability at least $c$ over
his random choices.
 \item $\forall x\notin L$ and for all $n$ provers  who share any $n$-partite state $\ket{\Psi}$ the interaction between V and the
provers results in the verifier accepting with probability at most $s$ over his random choices.
\end{itemize}
\end{definition}
Most of the time we consider only $2$-prover $1$-round protocols and omit the $(2,1)$.


To show our main result we will work with the following {\em gapped} instance of {\sc 3D-MATCHING}:

\begin{definition}\label{def:GM}
An instance of  $\eps$-\GM\, of size $n$ is given by three sets  $U,V,W$ with $|U|=|V|=|W|=n$, and a subset
$M\subset U\times V\times W$. For a positive instance there exist two bijections $\pi:U\rightarrow V$ and
$\sigma :U\rightarrow W$ such that
$$\forall u\in U\qquad (u,\pi(u),\sigma(u))\in M$$
\noindent  For a negative instance, for all bijections $\pi:U\rightarrow V$ and $\sigma :U\rightarrow W$, at
most a fraction $\eps$ of triples $(u,\pi(u),\sigma(u))$, for $u\in U$, are in $M$.
\end{definition}


\begin{fact}\label{constantdegree}
There exists constants $\Delta\in \N$ and $\eps>0$ such that the restriction of $\eps$-\GM\ to instances
where $M$ has outgoing degree bounded by $\Delta$ (for each $u \in U$ there  are neighborhoods $N_V(u)
\subset V$ and $N_W(u) \subset W$ such that $|N_V(u)|,|N_W(u)| \leq \Delta$ and if $(u,v,w) \in M$ then $v
\in N_V(u)$ and $w \in N_W(u)$) is still NP-complete.
\end{fact}

\begin{proof}
It is a direct consequence of the PCP theorem that there is a constant $\eps>0$ for which
$\eps-\textsc{gap-3SAT}$ is NP-complete \cite{papadimitriou:cc}. Applying the standard reduction from {\sc
3SAT} to {\sc 3DM} \cite{gareyjohnson} to \textsc{gap-3SAT} immediately yields the desired result. To give an
idea of parameter values, we obtain $\eps\simeq 1-1/8$ and $\Delta= 6$.
\end{proof}

\section{Proof idea and zero-error case}\label{sec:zero}

There is a generic classical \MIP\ protocol for \GM: the verifier picks a random vertex $u$ and sends it to
each of the two provers, asking them to apply bijections $\pi$ and $\sigma$. In the case of a positive
instance the provers send back $\pi(u)$ resp. $\sigma(u)$ and the verifier checks that $(u,\pi(u),\sigma(u))
\in M$. To enforce a bijection, the verifier performs another test with some probability: he picks random
vertices $u$ and $u'$ and asks both provers to apply $\pi$. He checks that the answers are the same if $u=u'$
and that the answers are different if $u\neq u'$. To have a constant probability of detecting cheating
provers, the verifier picks $u'$ among the neighbors of the neighbors of $u$. Since the degree of the
underlying graph is constant, the probability to detect a non-bijection is constant. For a negative instance
only a small fraction of $(u,\pi(u),\sigma(u))$ are in $M$ for any bijection, and hence the provers cannot
cheat.

The difficult part in giving a \QMS\ protocol for \GM\ is to show that entanglement does not help the provers
to coordinate their replies in order to cheat in a negative instance, i.e. to show reasonable {\em
soundness}. The idea is to use quantum messages and quantum tests, like the SWAP-test, to enforce an
(approximate) bijection from the provers.

In this section we first describe a QMIP$^*$ protocol for $3$-{\sc DM} and show its correctness in the case
of {\em zero-error}, i.e. under the assumption that the provers have to pass all the tests with probability
$1$. This allows us to present the basic ideas needed in Section \ref{sec:sound} to relax the soundness to
$1-2^{-O(n)}$.

\subsection{Description of the protocol}\label{sec:protocol}

The provers, called Alice and Bob, share some general entangled state $\ket{\Psi}$, which might depend on the
instance $x$ of \GM. The verifier V, who has a workspace of $O(\log n)$ qubits, sends simultaneously one
question to each prover, which consists of a single bit ($\pi$ or $\sigma$) and a register on $\log n$
qubits. We will use subscripts to indicate the registers sent to A and B and into which A and B will write
their answers, i.e. $\ket{\cdot}_A$ is send to Alice, she performs some operation on her space and the
register and sends it back, and similarly $\ket{\cdot}_B$ is sent to Bob. V begins by flipping two fair coins
with outcomes $\pi/\sigma$, and sends the result of the first coin flip to the first prover, and the result
of the second to the second prover. If both coins give the same result ($\pi,\pi$ or $\sigma,\sigma$) the
verifier does a set of tests that ensure that $\pi$ resp. $\sigma$ are bijections (\BIJ-Test $1$). Otherwise
the verifier tests if the instance of \GM\ is positive (\MA-Test 2). Note that in a part of Test 1 we use the
SWAP test \cite{bcww:fp}, that measures how similar two quantum states $\ket{\alpha}$ and $\ket{\beta}$ are.
Suppose $\ket{\alpha}$ and $\ket{\beta}$ are given in two separate registers. An ancillary qubit is prepared
in the state $\frac{1}{\sqrt{2}}(\ket{0}+\ket{1})$. This qubit controls a SWAP between the two registers, and
a Hadamard transform is applied to the ancillary qubit, which is then measured. The success probability, the
probability to measure $\ket{0}$, is given by $\frac{1}{2}(1+|\inp{\alpha}{\beta}|^2)$.

We denote elements of $U$ by $u$ and $u'$, elements of $V$ by $v$ and $v'$ and elements of $W$ by $w$ and
$w'$.

\paragraph{Test 1 (\BIJ)}
Let us assume that both coins gave $\pi$ (otherwise replace all $\pi$ with $\sigma$ and $v,v' \in V$ by $w,w'
\in W$). With probability $1/3$ the verifier prepares one of the following states, sends the corresponding
registers to A and B, receives their answers and performs a corresponding test:

\noindent {\bf a)} State: for a random $u \in U$
$$\frac{1}{\sqrt{2}}\left(\ket{0}\frac{1}{\sqrt{n}}\sum_{u'}\ket{u'}_A\ket{u}_B +  \ket{1}
\ket{u}_A\ket{u}_B \right).
$$
Test: This test incorporates three subtests:

1) If the first register is in state $\ket{1}$ the verifier checks that the answers of the provers are the
same. In other words he projects onto the space spanned by
$\{\ket{1}\ket{v}_A\ket{v}_B,\ket{0}\ket{v}_A\ket{v'}_B\,\, , v,v' \in V \}$, accepts iff the result is
positive and then controlled on the first register being $\ket{1}$ erases register $2$ by XORing register $3$
onto register $2$, such that register $2$ is in the state $\ket{0}_A$.

2) If the first register is in state $\ket{0}$, the verifier projects the second register onto
$\frac{1}{\sqrt{n}} \sum_v \ket{v}_A$, accepts iff the result is positive and then erases this register by
applying a unitary that maps $\frac{1}{\sqrt{n}} \sum_v \ket{v}_A$ to $\ket{0}_A$.

3) He measures the first register in the $\{\ket{+}$, $\ket{-}\}$ basis. If he gets $\ket{-}$, he rejects,
otherwise he accepts.

\noindent {\bf b)} Like {\bf a)} but with the registers $2$ and $3$ swapped.

\noindent {\bf c)} State: $$\frac{1}{n} \sum_{u,u'} \ket{u}\ket{u}_A\ket{u'}\ket{u'}_B$$ Test: Perform a
SWAP-test between registers 1,2 and 3,4. Accept if and only if it succeeds.

\paragraph{Test 2 (\MA)}
If the coins gave different results, then for a random $u \in U$ prepare state $\ket{u}\ket{u}_A\ket{u}_B$
and send register $2$ to Alice and $3$ to Bob. Receive their answers. Measure all registers in the
computational basis and get a triple $(u,v,w)$ (or $(u,w,v)$, depending on who got the $\pi$ and who got the
$\sigma$) as a result. Accept if $(u,v,w)\in M$ and reject otherwise.

\paragraph{Remarks:} Note that the \MA\ is completely classical.
The first part of the \BIJ, a)1) (and b)1)), simply checks that the provers give the same answer when confronted with
the same question. This part of the test is in fact entirely classical. As will become clear, the second part, a)2), is included only for convenience as it allows us to introduce a handy basis the zero-error case.
This test will be dropped in the general case. The third part, a)3) (resp. b)3)), serves to establish
that the provers indeed implement a bijection in some basis, that might depend on $u$. However it is part c) of the \BIJ, which is genuinely quantum, that allows us to show that there is a {\em global} basis in which the
prover's action is a bijection. It is this test that links our results to the $\delta$ in Conjecture
\ref{conj} in the non-zero-error case. We do not know if it is possible to find a classical test that would
establish this, but our attempts make us believe that it is unlikely and that we indeed need quantum messages
to establish the result.

\subsection{Zero-error proof}\label{sec:zero-errorproof}

First note that the verifier requires only space and time $O(\log n)$ for the execution of the protocol, if
he has access to his input through an oracle that given $u$ outputs all triples $(u,v,w) \in M$, of which there are a constant number. Moreover perfect completeness
($c=1$) follows trivially: for a positive instance of \GM\ there exist bijections $\pi:U\rightarrow V$ and
$\sigma:U\rightarrow W$ (from Def. \ref{def:GM}) such that if the provers apply the transformations
$\ket{u}\mapsto\ket{\pi(u)}$ and $\ket{u}\mapsto\ket{\sigma(u)}$ on their registers it is easy to check that
they are accepted with probability $1$ by the verifier.

We now show the converse: if two provers are accepted by the verifier with probability $1$ in the \BIJ
and with some constant probability in the \MA, then the instance of \GM\ is positive. More precisely
we show that if the provers pass the \BIJ, then their actions correspond to bijections (this will be made
precise below). Hence, if they also pass the \MA\ then there must be an approximate matching. At the beginning
of the protocol the joint state of A and B can be described as $\ket{\Psi}=\sum_{i\in I} \alpha_i
\ket{i}\ket{i}$ where $\{\ket{i}: i \in I \}$ is some orthonormal family (the Schmidt basis of A and B's
joint state including their private workspace) and $I$ can be arbitrarily large. Note that a priori there can
be several valid bijections $\pi_i$ and $\sigma_i$ such that $(u,\pi_i(u),\sigma_i(u)) \in M$ for all $u \in
U$. In particular the following is a perfectly valid action of A and B to pass the \MA: \[ \frac{1}{\sqrt{n}}
\sum_u \ket{u}\ket{u}_A \ket{u}_B \sum_{i \in I} \alpha_i \ket{i}\ket{i} \longrightarrow \frac{1}{\sqrt{n}}
\sum_{u,i} \alpha_i \ket{u}\ket{\pi_i(u)}_A \ket{\sigma_i(u)}_B (U_A\ket{i}) \otimes (V_B\ket{i}) \] for some
arbitrary unitary $U_A$ on A's system and $V_B$ on B's. Here, A and B use their entanglement as a shared coin
to chose one of the possible valid bijections. We will show that if they pass the \BIJ, this is the most
general thing they can do (up to local unitaries on their systems before answering V's questions). We need
some more notation to describe the action of Alice and Bob. Without loss of generality we assume that A's and
B's actions are unitary (by allowing them to add extra qubits to their workspace). Let $\textbf{A}^{\pi}$ and
$\textbf{A}^{\sigma}$ be the two unitaries that Alice applies to the question she receives and to her private
qubits (including the entanglement) before returning her answer, depending on the first bit she receives.
Similarly, Bob is described by $\textbf{B}^{\pi}$ and $\textbf{B}^{\sigma}$. Write the action of $\textbf{A}$
and $\textbf{B}$ (we often omit the $\pi$ and $\sigma$ superscripts when the context is clear) as
$$\ket{u}\ket{i}\mapsto \textbf{A}^{\pi} \ket{u}\ket{i}= \Sum{v}{} \ket{v}\ket{\ph^{\pi}(u,v,i)} \quad \quad
\quad \ket{u}\ket{i}\mapsto \textbf{B}^{\pi} \ket{u}\ket{i} = \Sum{v}{} \ket{v}\ket{\Psi^{\pi}(u,v,i)} $$ We
decompose $\textbf{A}$ into sub-matrices $A^{u,v}$ corresponding to $\{\ket{u}\}$ and $\{\ket{v}\}$ in this
definition. Similarly for $\textbf{B}$. $A^{u,v}$ is thus the matrix with column vectors
$\{\ket{\ph(u,v,i)},\, i\in I\}$ expressed in some basis $\{\ket{e_i}\}$, independant of $u$, which we will
define later, i.e. $A^{u,v}_{i,j}=\inp{e_i}{\ph(u,v,i)}$.  We would like to show that up to local unitaries
on the second system we have $\textbf{A}^{\pi} \ket{u}\ket{i} = \ket{\pi_i(u)} \ket{i}$, i.e.
$\ket{\ph^{\pi}(u,v,i)}=\ket{i}$ if $v=\pi_i(u)$ and zero otherwise. In what follows we will use the
following fact, which can be easily computed from the definitions. Let $D$ be the diagonal matrix having the
$\alpha_i$'s on its diagonal.

\begin{fact}
$\|\sum_{i \in I} \alpha_i \ \ket{\ph(u,v,i)}\ket{\Psi(u',v',i)}\|_2 =\|A^{u,v}D(B^{u',v'})^T\|_F$ where
$\|\cdot\|_2$ is the $L_2$ norm $\|\ket{v}\|_2^2=\inp{v}{v}$ and $\|\cdot \|_F$ is the Frobenius norm defined
as $\|A\|_F^2=\Tr(A^\dagger A)$.
\end{fact}

\begin{lemma}\label{lem:zero}
Assume the provers pass the \BIJ\ with probability $1$. Then there exist diagonal projector matrices
$P^{u,v}$ and $Q^{u,v}$ such that $\sum_u P^{u,v}=\sum_v P^{u,v}=I$ and $\sum_u Q^{u,v}=\sum_v Q^{u,v}=I$ and
unitary matrices $U_1$ and $V_1$ such that
$$ \forall (u,v)\in U\times V \quad \quad A^{u,v}=U_1P^{u,v}U_1^\dagger \quad \text{and}\quad B^{u,v}=V_1Q^{u,v}V_1^\dagger .$$
\end{lemma}
The fact that all $P^{u,v}$ are diagonal projectors together with the conditions $\sum_u P^{u,v}=\sum_v
P^{u,v}=I$ ensures that for a fixed $i$ and $u$ there is exactly one $v$ such that $P^{u,v}$ has a 1 in
position $i$ and vice-versa. This means that for fixed $i$, we can define a bijection $\pi_i$ by letting
$\pi_i(u)$ be the unique $v$ such that $(P^{u,v})_{i,i}=1$. In other words  if $\textbf{P}= U_1^\dagger
\textbf{A} U_1$, then $\textbf{P}\ket{u}\ket{i}=\sum_v \ket{v} P^{u,v} \ket{e_i}=\ket{\pi_i(u)}\ket{e_i}$.
$U_1$ is a local unitary on the prover's register only.

\begin{proof}
We begin with a claim summarizing the consequences of each of parts $a)$, $b)$ and $c)$ of the \BIJ.
\begin{claim}\label{claim:tests-ze}
As a consequence of Test 1, the following matrix relations hold for all  $u,u'\in U$ and $v,v'\in V$
\begin{subequations}\label{eq:tests-zeroerror}
    \begin{gather}
A^{u,v'}D(B^{u,v})^T=0 \quad if \, v' \neq v \label{eq:testaz-zeroerror}\\
  \label{eq:addb-zeroerror}  {A}^{u,v}D={A}^{u,v}D({B}^{u,v })^T=D({B}^{u,v })^T \\
A^{u,v}D(B^{u',v'})^T-A^{u',v'}D(B^{u,v})^T= 0\label{eq:test1c-zeroerror}
\end{gather}
\end{subequations}
\end{claim}

\begin{proof}
Let us first analyze part 1. of Test 1a). If the first qubit is in the state $\ket{1}$, the state of the
system after the provers have sent back their answers is
$$\sum_{v,v'}\ket{v}_A\ket{v'}_B\sum_{i} \alpha_i
\ket{\ph(u,v,i)}\ket{\Psi(u,v',i)}
$$
The probability to reject is given by the norm squared of the part of the state with $v \neq v'$, averaged
over all $u$, and hence we get
 \be \label{eq:testa}
 \frac{1}{n} \sum_{u,v,v':v \neq v'}\|\sum_{i} \alpha_i \ket{\ph(u,v,i)}\ket{\Psi(u,v',i)}\|_2^2=\frac{1}{n}
\sum_{u,v,v':v \neq v'}\|A^{u,v}D(B^{u,v'})^T\|_F^2=0
 \ee
which proves Eq. (\ref{eq:testaz-zeroerror}).

For part 2. of Test a), if the first qubit is in the state $\ket{0}$, the state of the system after the
provers have sent their answer is
$$\frac{1}{\sqrt{n}}\sum_{v'}\ket{v'}_A\sum_{v}\ket{v}_B\sum_{i} \alpha_i
\sum_{u'} \ket{\ph(u',v',i)}\ket{\Psi(u,v,i)}
$$
If provers pass part 2 of Test a) with probability $1$, the state must be a tensor product with
$\frac{1}{\sqrt{n}}\sum_{v'}\ket{v'}_A$ in the first register and hence the other registers must be
independent of $v'$. In other words $\ket{e_i}:=\sum_{u'} \ket{\ph(u',v',i)}$ is independent of $v'$. Note
that since Alice's transformation is unitary, it must be that the set of vectors $\{\ \frac{1}{\sqrt{n}}
\sum_{v'}\ket{v'} \Sum{u'}{} \ket{\ph(u',v',i)},\, i\in I\}$ are orthonormal, and hence the vectors
$\ket{e_i}$ also form an orthonormal basis. It is in this basis that we express the matrices $A^{u,v}$. Note
that in particular $\sum_u A^{u,v}=I$.
From part 2. of Test b) we similarly get a basis  $\ket{f_i}$.

In part 3. of Test a) the probability to measure $\ket{-}$ is given by the norm squared of the state
$$
\sum_{v} \ket{v}\Sum{i}{}\alpha_i \Big( \sum_{u'} \ket{\ph(u',v',i)}\ket{\Psi(u,v,i)}
-\ket{\ph(u,v,i)}\ket{\Psi(u,v,i)}\Big)
$$
averaged over all $u$. So we have for all $u,v$
 \be \label{eq:test3}
\|\Sum{i}{}\alpha_i \left(  \ket{e_i} -\ket{\ph(u,v,i)}\right)\ket{\Psi(u,v,i)}\|_2^2=\| (I  -
A^{u,v})D(B^{u,v})^T\|_F^2=0,
 \ee
i.e.  $ D(B^{u,v})^T=A^{u,v}D(B^{u,v})^T$. From part 3. of Test b), similarly $A^{u,v}D=A^{u,v}D(B^{u,v})^T$,
which combined give Eq. (\ref{eq:addb-zeroerror}).

We finally exploit Test 1(c). The SWAP-test succeeds with probability 1 if the norm of the state
$$
\frac{1}{n}\sum_{u,u',v,v'}\ket{u}\ket{v}_A\ket{u'}\ket{v'}_B\sum_i \alpha_i
\left(\ket{\ph(u,v,i)}\ket{\Psi(u',v',i)}-\ket{\ph(u',v',i)}\ket{\Psi(u,v,i)}\right)
$$
is zero. This immediately implies Eq. (\ref{eq:test1c-zeroerror}).
\end{proof}
\begin{claim}\label{claim:adaggera-ze}
     The matrices $A^{u,v}$ are projectors. More precisely,
\be\label{adaggera-ze}
 \forall u,v\in U\times V\quad\quad A^{u,v}=(A^{u,v})^\dagger A^{u,v}
\ee
 \end{claim}
\begin{proof}
 With the notation that $(X)_j$ is the $j$th column of a matrix $X$, write
 \begin{align}\label{eq:adagger0}
\ket{v}\otimes ({A}^{u,v} D)_j=\ket{v} \otimes ( {A}^{u,v}D(B^{u,v})^T)_j&=\Sum{v'}{}
\ket{v'}\otimes(A^{u,v'} D(B^{u,v})^T)_j \nonumber \\&= \Sum{i}{}\alpha_i {B}_{j,i}^{u,v}
\Sum{v'}{}\ket{v'}\otimes\ket{\ph(u,v',i)}
 \end{align}
where we used (\ref{eq:testa}). Since $\{\Sum{v'}{} \ket{v'}\otimes \ket{\ph(u,v',i)},\, i\in I\}$ are
orthonormal, we get $ \alpha_j \inp{\ph(u,v,i)}{\ph(u,v,j)}-\alpha_i {B}_{j,i}^{u,v} =0$, i.e.
$(A^{u,v})^\dagger A^{u,v}D=D(B^{u,v})^T$, which, using (\ref{eq:addb-zeroerror}), finally gives Eq.
(\ref{adaggera-ze}). So ${A}^{u,v}$ is a diagonalizable matrix with eigenvalues in $\{ 0,1\}$.
\end{proof}

Combining Eqs. (\ref{eq:addb-zeroerror}) and (\ref{eq:test1c-zeroerror}) we have that
$A^{u,v}A^{u',v'}-A^{u',v'}A^{u,v} = 0$ for all $u,u',v,v'$, i.e. the matrices $A^{u,v}$ are mutually
commuting, and thus simultaneously diagonalizable. Let $U_1$ be the diagonalization matrix. We have

$$\forall u,v\in U\times V \qquad A^{u,v} = U_1 P^{u,v} U_1^\dagger \quad\text{and}\quad B^{u,v}=V_1 Q^{u,v}
V_1^\dagger$$ where $P$ and $Q$ are diagonal matrices with eigenvalues $0,1$. Finally, since the family
$\{\Sum{v}{} \ket{v}\otimes \ket{\ph(u,v,i)},\, i\in I\}$ is orthonormal, we have
$\Sum{v}{}\inp{\ph(u,v,i)}{\ph(u,v,j)}=\delta_{i,j}$ and hence $\Sum{v}{} (A^{u,v})^\dagger A^{u,v}=\Sum{v}{}
{A}^{u,v} = I$.

\end{proof}

\begin{lemma}
For a negative instance of $\eta$-\GM\, if the provers pass the \BIJ\ with probability 1 they will fail
the \MA\ with probability at least $1-\eta$.
\end{lemma}
Without loss of generality assume the verifier sends $\pi$ to Alice and $\sigma$ to Bob. From Lemma
\ref{lem:zero} we know that Alice implements $\textbf{A}^{\pi}=U_1 \textbf{P}^{\pi} U_1^\dagger$ and Bob
$\textbf{B}^{\sigma}=V_1 \textbf{Q}^{\sigma} V_1^\dagger$ where
$\textbf{P}^{\pi}\ket{u}\ket{i}=\ket{\pi_i(u)}\ket{e_i}$ and
$\textbf{Q}^{\sigma}\ket{u}\ket{i}=\ket{\sigma_i(u)}\ket{f_i}$. Hence the state the verifier receives is
 \be \label{eq:goodstate}
\ket{\alpha(u)}:=\ket{u}\textbf{A}^{\pi}\ket{u}_A\textbf{B}^{\sigma}\ket{u}_B =\ket{u}\Sum{j,k}{}
\ket{\pi_j(u)}_A\ket{\sigma_k(u)}_B \left(U_1 D V_1^\dagger \right)_{j,k} U_1 \ket{e_j} \otimes  V_1
\ket{f_k}
 \ee

V measures the triple $(u,\pi_j(u),\sigma_k(u))$ with probability $\left|(U_1 D V_1)_{j,k}^\dagger \right|^2$
which is \emph{independent of $u$}. For a negative instance we know that for any bijection $\pi_j$ and
$\sigma_k$ for a fraction of at least $1-\eta$ of the $u$, $(u,\pi_j(u),\sigma_k(u))\notin M$ and so the
provers fail Test 2 with probability at least $1-\eta$.

Note that the proof still works if the state that the verifier receives is not exactly equal to the state in
(\ref{eq:goodstate}).

\begin{claim}\label{claim:fidelity}
Assume the state $\ket{\alpha'(u)}$ of the verifier after receiving the provers registers in the \MA\ is such
that $\frac{1}{n} \sum_{u} |\inp{\alpha(u)}{\alpha'(u)}|^2 \leq \delta$, then in the case of a negative
instance of $\eta$-\GM\ they will fail the \MA\ with probability at least $1-\eta-\delta$.
\end{claim}
This follows because the two density matrices $\rho=\frac{1}{n}\sum_u
\ketbra{u}{u}\otimes\ketbra{\alpha(u)}{\alpha(u)}$ and $\rho'=\frac{1}{n}\sum_u
\ketbra{u}{u}\otimes\ketbra{\alpha'(u)}{\alpha'(u)}$ have fidelity $1-\delta$ and hence the probability to
accept when given $\rho$ differs from the probability to accept when given $\rho'$ by at most $\delta$.

\section{Decreasing soundness}\label{sec:sound}

In this section we prove Theorem \ref{thm:main} and Corollary \ref{cor:main}. To deal with error, we begin by
slightly modifying the protocol introduced in \ref{sec:protocol}. We only make changes to parts a) and b) of
the \BIJ.

Part 1. of test a) (and b)) is modified in the following way: after receiving the prover's answers, we will
flip a fair coin and, if the result is $0$, then we will project onto the space spanned by the vectors
$\{\ket{1}\ket{v}\ket{v},\ket{0}\ket{v'}\ket{v}\,\, , v' \in V, v \in N_V(u) \}$ and accept if and only if we
get a positive result. If the result of the coin flip was zero, we project onto
$\{\ket{1}\ket{v}\ket{v},\ket{0}\ket{v'}\ket{v}\,\, , v,v' \in V\}$ as in the original test, and proceed
directly to part $3$ of the test. We thus completely drop part 2 of Test a (and b), which was used in the
zero-error case to introduce the basis $\ket{e_i}$. Since we do not want to deal with approximately
orthonormal bases, we will replace it by a perfectly orthonormal basis $\ket{\tilde{e}_i}$, with the caveat
that it is inside a larger Hilbert space. All the other tests remain the same.

As in the zero-error proof, the key lemma  states that provers who pass the \BIJ\ with probability 1-$\eps$ must
apply approximate bijections. More precisely, we prove the following

\begin{lemma}\label{lem:nonzero}
Assume the provers pass the \BIJ\ with probability $1-\eps$. Then there exist a constant $C>0$ and
diagonal projectors $P^{u,v}$ and $Q^{u,v}$ such that $\sum_u P^{u,v}=\sum_v P^{u,v}=I$ and $\sum_u
Q^{u,v}=\sum_v Q^{u,v}=I$ and unitary matrices $U_1$ and $V_1$ such that
$$ \frac{1}{n} \sum_{u;v}\|(A^{u,v}-U_1P^{u,v}U_1^\dagger)D \|_F^2 \leq C^n\eps \quad\text{and} \quad
\frac{1}{n} \sum_{u;v}\| (B^{u,v}-V_1Q^{u,v}V_1^\dagger)D \|_F^2 \leq C^n\eps.$$
\end{lemma}

To conclude Theorem \ref{thm:main} from this lemma, note that, as in Section \ref{sec:zero-errorproof}, the
verifier uses space $O(\log n)$. Perfect completeness follows again trivially. Let $\eps$ be the constant
from Fact \ref{constantdegree}. Suppose that the two provers pass the \BIJ\ with probability $1-C^{-n}\eps/2$,
and the \MA\ with constant probability $1-\eps/2$. Then, Lemma \ref{lem:nonzero} together with Claim
\ref{claim:fidelity} imply that the instance of \GM\ must be positive. This proves that our protocol has
soundness $1-C^{-n}$. To conclude Corollary \ref{cor:main}, observe that the bottleneck to decreased
soundness comes from Test 1c) and Lemma \ref{lem:diagonal}. From the proof of Lemma \ref{lem:nonzero} it
follows that if Conjecture \ref{conj} is true for some $\delta(m,\eps)$, then Lemma \ref{lem:nonzero} is true
when $C^n \eps$ is replaced by $\delta(C'n,C''\eps)$ for some constants $C',C''>0$.

We will use the following easy facts in our proof:

\begin{fact}\label{fact:delta}
(a) Let $\|\cdot \|_{op}$ be the operator norm (largest singular value). If $\|A\|_{op} \leq 1$ then
$\|AB\|_F \leq \|B\|_F$. (b) (Triangle inequality) For a constant number of matrices $X_1,\ldots,X_{\Delta}$
we have $\|\sum_{i=1}^\Delta X_i\|_F^2 \leq (\sum_{i=1}^\Delta \|X_i\|_F)^2 \leq \Delta^2 \max_i(
\|X_i\|_F^2)$
\end{fact}

\begin{fact}\label{fact:unitary}
Let $U=\left(\begin{array}{cc}\tilde{U}_0 & \tilde{U_1}\\\tilde{U_2} & \tilde{U}_3\end{array}\right)$ be a
unitary matrix such that $\|\tilde{U}_2D\|_F^2=O(\eps)$ and $\tilde{U}_0$ is a square matrix. Then there
exists a unitary matrix $U_0$ such that $\|({U}_0-\tilde{U}_0)D\|_F^2=O(\eps)$.
\end{fact}

\begin{proof}
Since $U^\dagger U = I$ we have that $ \|(\tilde{U}_0^\dagger \tilde{U}_0 - I)D \|_F^2 = O(\eps)$. Let
$\tilde{U}_0=P Z Q^\dagger$ be the singular value decomposition of $\tilde{U}_0$ with singular values
$\lambda_i \geq 0$ and define ${U}_0= P  Q^\dagger$ (which as a product of unitaries is unitary). Then
$\tilde{U}_0^\dagger \tilde{U}_0 = Q Z^\dagger Z Q^\dagger$, and we get $\|(Z^\dagger Z-I)Q^\dagger D\|_F^2 =
\sum_{i,j} |(\lambda_i^2-1)\bar{Q}_{j,i}\alpha_j \|^2= O(\eps).$ Since $|\lambda_i-1| \leq
|\lambda_i-1|(\lambda_i+1) = |\lambda_i^2-1|$, we finally have
$$\| (U_0 - \tilde{U}_0)D\|_F^2 =\|(Z-I)Q^\dagger)D\|_F^2= \sum_{i,j} |(\lambda_i-1)\bar{Q}_{j,i}\alpha_j |^2 \leq
\sum_{i,j} |(\lambda_i^2-1)\bar{Q}_{j,i}\alpha_j |^2 = O(\eps).$$
\end{proof}

\paragraph{Notations:} Let us start by describing the matrix notations we use in the proof of Lemma
\ref{lem:nonzero}. As in Section \ref{sec:zero}, $A^{u,v}$ is the square matrix with columns
$\{\ket{\ph(u,v,i)},\, i\in I\}$ expressed in a basis $\ket{e_i}$ which will be defined later. Let
$\ket{\tilde{e}_i}:= \frac{1}{\sqrt{n}} \sum_{v'} \ket{v'}\sum_{u'}\ket{\ph(u',v',i)}$. The family
$\{\ket{\tilde{e}_i},\,i\in I\}$ is orthonormal as an immediate consequence of the prover's unitarity. This
family is included in the Hilbert space $\tilde{\mathcal{H}}$ spanned by all vectors of the form
$\ket{v}\ket{i}$ for $v\in V$ and $i\in I$. We complete this family to a basis $\{\ket{\tilde{e}_i},\,i\in
J\}$ of $\tilde{\mathcal{H}}$, where $|J|=|I|\cdot|V|$. Letting
$\ket{\tilde{\ph}(u,v,i)}=\frac{1}{\sqrt{n}}\sum_{v'} \ket{v'} \ket{\ph(u,v,i)}$, $\tilde{A}^{u,v}$ is the
rectangular matrix with column vectors $\ket{\tilde{\ph}(u,v,i)}$ expressed in the basis $\ket{\tilde{e}_i}$.
Define $\tilde{A'}^{u,v}$ as the matrix equal to $\tilde{A}^{u,v}$ with all rows below the $|I|$th row set to
$0$. Finally $\tilde{I}_A$ is the matrix of same dimensions as $\tilde{A}$ formed by an $|I|\times |I|$ block
equal to the identity matrix over a rectangular block of zeroes, and $\hat{A}^{u,v}=\tilde{I}_A^T
\tilde{A}^{u,v}$ is the upper block of $\tilde{A}$. Matrices $B^{u,v}$, $\tilde{B}^{u,v}$,
$\tilde{B'}^{u,v}$, $\hat{B}^{u,v}$ and $\tilde{I}_B$ are defined in the same way for the vectors
$\ket{\Psi(u,v,i)}$, in bases $\ket{f_i}$ and $\ket{\tilde{f}_i}$. The relations between all these matrices
will be given in (\ref{eq:utildea}) and (\ref{eq:ahata}).

\begin{proof}[Proof of Lemma \ref{lem:nonzero}:]
The idea is to follow the lines of the proof of  Lemma \ref{lem:zero} and to prove  {\em approximate}
versions of Claim  \ref{claim:tests-ze} (Claim \ref{claim:tests}) and Claim \ref{claim:adaggera-ze} (Claim
\ref{claim:adaggera}).

\begin{claim}\label{claim:tests}
The following matrix relations hold as a consequence of Test 1
\begin{subequations}\label{eq:tests}
    \begin{gather}
\frac{1}{n}\sum_{u}\bigg(\mathop{\sum_{v \in N_V(u)}}_{v':v' \neq v }\|A^{u,v'}D(B^{u,v})^T\|_F^2 +
\sum_{v \notin N_V(u); v'}\|A^{u,v'}D(B^{u,v})^T\|_F^2\bigg)=O(\eps) \label{eq:testaz}\\
  \label{eq:addb}  \frac{1}{n}\sum_{u,v}\| \tilde{A}^{u,v}D
\tilde{I}_B^T-\tilde{A}^{u,v}D(\tilde{B}^{u,v })^T\|_F^2=O(\eps)\quad   \frac{1}{n}\sum_{u;v\in N_V(u)} \|
\tilde{A}^{u,v}D \tilde{I}_B^T-
\tilde{I}_AD(\tilde{B}^{u,v })^T\|_F^2=O(\eps) \\
\frac{1}{n^2} \sum_{u,u';v \in N_V(u);v' \in N_V(u')} \|(A^{u,v}D(B^{u',v'})^T-A^{u',v'}D(B^{u,v})^T\|_F^2 =
O(\eps)\label{eq:test1c}
\end{gather}
\end{subequations}
\end{claim}

\begin{proof}
Since we assume that the provers pass Test $1$ with probability at least $1-\eps$, they must pass each of the
Tests 1a, 1b and 1c with probability at least $1-3\eps$.

We first study the consequences of Test 1a.  The verifier flips a fair coin. The provers must have a success
probability of at least $1-6\eps$ in any of the two cases. If the verifier got a $0$, Eq. (\ref{eq:testa})
becomes
 \be
\frac{1}{n}\sum_{u}\bigg(\mathop{\sum_{v \in N_V(u)}}_{v':v' \neq v }\|A^{u,v'}D(B^{u,v})^T\|_F^2 +\sum_{v
\notin N_V(u); v'}\|A^{u,v'}D(B^{u,v})^T\|_F^2\bigg)\leq 6\eps \nonumber
 \ee
 \noindent which gives (\ref{eq:testaz}).
 If the verifier's coin flip resulted in a $1$, assuming the provers pass the projection test in part 1,
with the convention that $\ket{0}_A=\frac{1}{\sqrt{n}}\sum_{v'}\ket{v'}_A$, the  state is projected onto
$$
\frac{1}{\sqrt{n}}\sum_{v ,v'} \ket{v'}_A \ket{v}_B \left(N_0
\ket{0}\sum_{i}\alpha_i\sum_{u'}\ket{\ph(u',v',i)}\ket{\Psi(u,v,i)}+ N_1 \ket{1}
\sum_{i}\alpha_i\ket{\ph(u,v,i)}\ket{\Psi(u,v,i)}\right)
 $$
 where $N_0$ and $N_1$ are normalization factors, $N_0,N_1 \geq 1/\sqrt{1-6\eps}$. In the following we will
not write these renormalisation factors with the understanding that the corresponding norms change by at most
factors of $1 \pm 6 \eps <2$, and we will write $O(\eps)$ for $c \cdot \eps$ where $c>0$ is some constant
independent of $n$. In part 3, the probability of measuring $\ket{-}$ is given by the (averaged over $u$)
norm square of
$$ \sum_{v} \ket{v}_B \sum_{i} \alpha_i \frac{1}{\sqrt{n}} \sum_{v'}
\ket{v'}_A \left( \sum_{u'}\ket{\ph(u',v',i)}\ket{\Psi(u,v,i)} - \ket{\ph(u,v,i)}\ket{\Psi(u,v,i)}\right).
$$
The norm inequality above can be rewritten in terms of the matrices $\tilde{A}^{u,v}$ similarly to
Eq.~(\ref{eq:test3})
 $$
\frac{1}{n}\sum_{u;v} \|\sum_{i} \alpha_i \left(\ket{\tilde{e}_i}- \ket{\tilde{\ph}(u,v,i)}\right)
\ket{\Psi(u,v,i)}\|_2^2=\frac{1}{n}\sum_{u;v} \| (\tilde{I}_A  - \tilde{A}^{u,v})D(\tilde{B}^{u,v})^T\|_F^2
=O(\eps), \nonumber
 $$
giving the first part of Eq. (\ref{eq:addb}). We obtain a symmetrical relation for matrices $\tilde{B}$ from
Test 2b). We combine them, using the triangle inequality and summing over $v\in N_V(u)$ only, to obtain the
second part of Eq. (\ref{eq:addb}). Finally, (\ref{eq:test1c}) follows directly from succeeding Test 1c) with
probability at least $1-3\eps$.

 \end{proof}

 \begin{claim}\label{claim:adaggera}
     The matrices $A^{u,v}$ are almost projector matrices. More precisely,
\be\label{adaggeraerror}
 \frac{1}{n}\sum_{u;v\in N_V(u)} \|(A^{u,v}-(A^{u,v})^\dagger A^{u,v})D\|_F^2=O(\eps).
\ee
 \end{claim}

 \begin{proof}
Note that the matrix $\tilde{A}D\tilde{I}_B^T$ has zero columns starting with the $|I|+1$st column and  the
matrix $\tilde{I}_AD\tilde{B}^T$  has zero rows starting with the $|I|+1$st row. Then the first part of
(\ref{eq:addb}) implies that \be\label{atildeprime}
 \frac{1}{n}\sum_{u;v \in N(u)} \|(\tilde{A}^{u,v}-\tilde{A}'^{u,v})D\|^2_F = O(\eps)
\ee
 and similarly for $\tilde{B'}$.

Let $\ket{\tilde{i}}=\frac{1}{\sqrt{n}}\sum_v\ket{v}\ket{i}\in\tilde{\mathcal{H}}$. Complete to a basis
$\{\ket{\tilde{i}},\,i\in J\}$ of $\tilde{\mathcal{H}}$. Let $U$ be the unitary that maps $\ket{\tilde{e}_i}$
to $\ket{\tilde{i}}$. Then $U\tilde{A}$ is a rectangular matrix consisting of a block equal to the original
$A$ matrix over a block of zeroes. This can be restated as $U\tilde{A}=\tilde{I}_A A $. Relation
(\ref{atildeprime}) can then be rewritten as \be\label{eq:utildea} \frac{1}{n}\sum_{u;v \in N(u)}
\|(\tilde{I}_A A^{u,v}-U\tilde{A'}^{u,v})D \|_F^2 = O(\eps)
 \ee
We now proceed similarly to the proof of (\ref{adaggera-ze}). We have $(\tilde{A}^{u,v'} D
(\tilde{B}^{u,v})^T)_j= \sum_i \alpha_i \tilde{B}^{u,v}_{j,i} \ket{\tilde{\ph}(u,v',i)}$ and
$$
\sum_i \alpha_i \tilde{B}^{u,v}_{j,i} \sum_{v'} \ket{v'}\otimes \ket{\tilde{\ph}(u,v',i)} = \ket{v}\otimes
\big(\tilde{A}^{u,v} D (\tilde{B}^{u,v})^T\big)_j+\sum_{v'\neq v} \ket{v'} \otimes \big(\tilde{A}^{u,v'} D
(\tilde{B}^{u,v})^T\big)_j
$$
Since the $\{\sum_{v}\ket{v}\otimes \ket{\tilde{\ph}(u,v,i)},\, i\in I\}$ are orthonormal, summing over
$v,i,j$ and averaging over $u$, using (\ref{eq:testaz}), (\ref{eq:addb}) and
$\inp{\tilde{\ph}(u,v,i)}{\tilde{\ph}(u,v,j)}=\inp{\ph(u,v,i)}{\ph(u,v,j)}$, this implies
 $$ \frac{1}{n}\sum_{u;v\in N_V(u)} \sum_{i,j\in I} |\alpha_i
\tilde{B}_{j,i}^{u,v} - \alpha_j\inp{{\ph}(u,v,i)}{{\ph}(u,v,j)}|^2 = O(\eps)
 $$
 so that
$\frac{1}{n}\sum_{u;v\in N_V(u)} \|D(\hat{B}^{u,v})^T - ({A}^{u,v})^\dagger {A}^{u,v}D\|_F^2 = O(\eps)$,
which using (\ref{eq:addb}) implies that \be\label{eq:atildedaggeraerror} \frac{1}{n}\sum_{u;v\in N_V(u)}
\|(\hat{A}^{u,v} - ({A}^{u,v})^\dagger {A}^{u,v})D\|_F^2 = O(\eps). \ee
 Let $S^{u,v}= ({A}^{u,v})^\dagger {A}^{u,v}$ be the square matrix with
coefficients $S^{u,v}_{i,j} = \inp{\ph(u,v,i)}{\ph(u,v,j)}$. We now show that
$\frac{1}{n}\sum_u\|(I-\sum_{v\in N_v(u)} S^{u,v})D\|_F^2=O(\eps)$. Considering first only the contribution
of the diagonal entries, we get
 \be
 \frac{1}{n}\sum_{u,i} \left(\left(1-\sum_{v\in N_V(u)} \|\ket{\ph(u,v,i)}\|^2\right)\alpha_i\right)^2 \leq
 \frac{1}{n}\sum_{u,i} \alpha_i^2  \left(1-\sum_{v\in N_V(u)} \|\ket{\ph(u,v,i)}\|^2\right) = O(\eps).
 \label{eq:sums}
 \ee
For the first inequality we use $\sum_{v} \|\ph(u,v,i)\|^2=1$, so that $0 \leq 1-\sum_{v\in N_v(u)}
\|\ph(u,v,i)\|^2 \leq 1$. Now combine (\ref{eq:testaz}) with (\ref{eq:addb}) to get
$\frac{1}{n}\sum_{u;v\notin N_V(u)} \|\tilde{A}^{u,v} D (\tilde{I}_B)^T\|_F^2=O(\eps)$ , which implies that
\\ $\frac{1}{n}\sum_{u;v\notin N_V(u),i}\alpha_i^2\|\ket{\ph(u,v,i)}\|^2 = O(\eps)$ (since
$\|\ket{\ph(u,v,i)}\|=\|\ket{\tilde{\ph}(u,v,i)}\|$). As $\sum_{v,i}\alpha_i^2\|\ket{\ph(u,v,i)}\|^2 =
1=\sum_i \alpha_i^2$, we get the second inequality in (\ref{eq:sums}).

As $\sum_v \ket{v}\ket{\ph(u,v,i)}$ is an orthonormal family over $i$, we have that for all $u$, $\sum_v
S^{u,v} = I$. All $S^{u,v}$ being positive matrices, $I-\sum_{v\in N_V(u)} S^{u,v}$ is also positive, write
it as $Y^\dagger Y$. Then the diagonal coefficients of $I-\sum_{v\in N_V(u)} S^{u,v}$ are the norms of the
column vectors of $Y$, so $\|YD\|_F^2 = O(\eps)$. Moreover, since $Y^\dagger Y\leq I$, $Y$ has operator norm
less than $1$. This implies that $\|Y^\dagger YD\|_F^2 = O(\eps)$, yielding the desired inequality.

Summing over $v$ and using Fact \ref{fact:delta} together with (\ref{atildeprime}),
(\ref{eq:atildedaggeraerror}), we get
$$
\frac{1}{n}\sum_{u}\, \|\big(\sum_{v\in N_V(u)} \tilde{A}^{u,v} - \sum_{v\in N_V(u)} \tilde{I}_A S^{u,v})D
\|_F^2 = O(\Delta\cdot\eps)=O(\eps)
$$
so, by (\ref{eq:sums}), since $U\tilde{A}=\tilde{I}_A A$, we get that $\frac{1}{n}\sum_{u}\,\| ( \tilde{I}_A
\sum_{v\in N_V(u)}A^{u,v}-U\tilde{I}_A)D  \|_F^2 = O(\eps)$. Let $\tilde{U}_0$ be the upper left block of $U$
and $\tilde{U}_2$ its lower left block. From the definition of $\tilde{I}_A$, this implies that
$\|\tilde{U}_2 D\|_F^2 = O(\eps)$ and $\frac{1}{n}\sum_u\|(\tilde{U}_0 -\sum_{v\in N_V(u)} A^{u,v})D \|_F^2 =
O(\eps)$. From Fact \ref{fact:unitary} we get a unitary $U_0$ such that $\|(U_0-\tilde{U}_0)D\|_F^2=O(\eps)$
and hence $\|({U}_0 -\sum_{v\in N_V(u)} A^{u,v})D \|_F^2 = O(\eps)$. We now choose the basis $\ket{e_i}$ in
which matrices $A^{u,v}$ are expressed to be the basis defined by $U_0^\dagger$ as
$\ket{e_i}=U_0^\dagger\ket{i}$. Equation (\ref{eq:utildea}) becomes
 \be\label{eq:ahata}
\frac{1}{n}\sum_{u; v\in N_V(u)}\| (\hat{A}^{u,v}-A^{u,v})D\|_F^2 = O(\eps)
 \ee
which, together with (\ref{atildeprime}), provides the link between matrices $A$, $\hat{A}$ and $\tilde{A}$.
We also have that $\frac{1}{n}\sum_{u} \|(I-\sum_{v\in N_V(u)}A^{u,v})D\|_F^2 = O(\eps)$, and, combining
(\ref{eq:atildedaggeraerror}) and (\ref{eq:ahata}) proves the claim.
\end{proof}

\begin{claim}\label{claim:projectors}
    There exist projectors $P^{u,v}$ such that
    \begin{subequations}\label{eq:diagonal}
        \begin{gather}
\frac{1}{n} \sum_{u;v \in N_V(u)} \|\left({A}^{u,v} - P^{u,v} \right)D\|_F^2 =O(\eps)
\label{eq:diagonalp}\\
\label{eq:diagonalq}\frac{1}{n^2}\sum_{u,u';v \in N_V(u);v' \in N_V(u')}
\|(P^{u,v}P^{u',v'}-P^{u',v'}P^{u,v})D\|_F^2 = O(\eps)
\end{gather}
\end{subequations}
\end{claim}

\begin{proof}
Claim \ref{claim:adaggera} implies that on average ${A}^\dagger A$ (and hence $A$) has eigenvalues close to
$0$ or $1$. More precisely, combining (\ref{eq:addb}) and (\ref{adaggeraerror}) with the triangle inequality,
$\frac{1}{n}\sum_{u,v\in N_V(u)} \|D(B^{u,v})^T - (A^{u,v})^\dagger A^{u,v}D\|_F^2=O(\eps)$. So
$\frac{1}{n}\sum_{u,v\in N_V(u)} \|S^{u,v}D(B^{u,v})^T - (S^{u,v})^2D\|_F^2=O(\eps)$, since $S$ has operator
norm less then $1$. Using (\ref{eq:addb}) to replace $SDB^T$ by $SD$, we finally get
 \be\label{eq:ss2}
\frac{1}{n}\sum_{u,v\in N_V(u)} \|(S^{u,v}-(S^{u,v})^2)D\|_F^2=O(\eps)
 \ee
Diagonalize $S^{u,v}$ as $S=U^{u,v}Z^{u,v}(U^{u,v})^\dagger $, where $Z$ is diagonal and let
$\lambda^{u,v}_i$ be its eigenvalues. Then (\ref{eq:ss2}) is rewritten as
$$\frac{1}{n}\sum_{u,v\in N_V(u)} \sum_{i,j} |(\lambda^{u,v}_i-(\lambda^{u,v}_i)^2)\bar{U}^{u,v}_{j,i}\alpha_j |^2 =O(\eps)$$
The $\lambda_i$ are such that $0\leq\lambda_i\leq 1$. Let $\mu_i$ be the nearest integer to $\lambda_i$. It
is easy to check that $|\lambda_i-\mu_i|\leq 2\lambda_i(1-\lambda_i)$, so
 $$\frac{1}{n}\sum_{u,v\in N_V(u)}\sum_{i,j} |(\lambda^{u,v}_i-\mu^{u,v}_i)\bar{U}^{u,v}_{j,i}\alpha_j |^2
 \leq 4 \frac{1}{n}\sum_{u,v\in N_V(u)} \sum_{i,j} |(\lambda^{u,v}_i-(\lambda^{u,v}_i)^2)\bar{U}^{u,v}_{j,i}\alpha_j |^2=O(\eps)$$
Let $P'^{u,v}$ be the diagonal matrix with entries $\mu^{u,v}_i$ if $v\in N_V(u)$, and $P'^{u,v}=0$ if
$v\notin N_V(u)$. Let $P^{u,v}:=U^{u,v}P'^{u,v} (U^{u,v})^{\dagger}$. Then
$$\frac{1}{n} \sum_{u;v \in N_V(u)} \|\left({S}^{u,v} - U^{u,v}
P'^{u,v} (U^{u,v})^{\dagger}\right)D\|_F^2 =O(\eps).$$ By Claim \ref{claim:adaggera}, this implies Eq.
(\ref{eq:diagonalp}).

From (\ref{eq:test1c}), using successively (\ref{eq:diagonalp}), (\ref{eq:addb}) and again
(\ref{eq:diagonalp}) together with the triangle inequality, since the projectors $P$ have operator norm
bounded by $1$, we get Eq. (\ref{eq:diagonalq}).
\end{proof}

By Markov's inequality Eq. (\ref{eq:diagonalq}) implies that for a subset $U' \subseteq U$ of size
$(1-O(\eps))|U|$  we have that $\|(P^{u,v}P^{u',v'}-P^{u',v'}P^{u,v})D\|_F^2=O(\eps)$ for $u,u' \in U'$. This
allows us to apply the following lemma, proving Conjecture \ref{conj} for $\delta=2^{O(n)}\eps$.

\begin{lemma}\label{lem:diagonal}
Assume that projectors $P_1,\ldots ,P_m$ are such that $\forall i, j$ we have $\|(P_iP_j-P_jP_i)D\|^2_F \leq
\eps$. Then there exist diagonal projectors $Q_1,\ldots , Q_m$, and a unitary matrix $U$, such that $\forall
i$ $\|(P_i-UQ_iU^\dagger)D\|_F^2 \leq c^n \eps$ for some constant $c$.
\end{lemma}
\begin{proof}
The proof is by brute force successive diagonalization. Choose a basis in which $P_1$ is diagonal and has
first a block of $1$s on the diagonal, followed by $0$s; this defines four blocks. Because of the commutation
relations we have that in this basis for all other $P_i$ the sum of the norms squared of the upper right and
lower left blocks is bounded by $\eps$. Set these blocks to $0$ in each $P_i$, apply a unitary that
diagonalizes the upper left and lower right blocks, round the eigenvalues to the closest integer ($0$ or
$1$), and apply the inverse of this unitary. After this first round we are left with new projectors
$P^1_2,\ldots ,P^1_m$ which are block-diagonal in a common block structure, with the two off-diagonal blocks
being $0$. Moreover, because of the cutting and rounding, the norms of the commutators of the new matrices
will be bounded by $c\eps$ for some constant $c$. They all commute exactly with $P_1$. $P_1$ will not be
changed any more.

In the next round choose a (block-diagonal) basis in which $P^1_2$ is diagonal such that inside the two
blocks defined by $P_1$ we first have a run of $1$s on the diagonal, followed by $0$s. Note that $P_1$ stays
diagonal in this basis, since it was either the identity or zero on each of the two blocks we are now
modifying. For the remaining projectors ($P^1_3,P^1_4,\ldots$) set the four resulting off-diagonal
sub-blocks, which have norm at most $c \eps$, to $0$ and re-round the eigenvalues as before. The resulting
projectors commute with $P_1$ and $P^1_2$ and the norm of their pairwise commutators is now bounded by $c^2
\eps$. Proceed in this way one by one with the remaining projectors. Each time the norms of the commutators
are at most multiplied by $c$. This gives the desired result.
\end{proof}

Applying Lemma \ref{lem:diagonal} to the $P'^{u,v}$, we get a set of commuting projectors $Q^{u,v}$ that are
simultaneously diagonalizable, and close to the $P'^{u,v}$ in Frobenius norm. To complete the proof of Lemma
\ref{lem:nonzero} it remains to prove that we can slightly modify these projectors so that they sum to the
identity on both $u$ and $v$. Recall that we proved that $\frac{1}{n}\sum_u \|(\sum_{v\in N_V(u)} S^{u,v} -
I)D\|_F^2 = O(\eps)$. From Claims \ref{claim:adaggera} and \ref{claim:projectors}), we get $\frac{1}{n}\sum_u
\|(\sum_{v\in N_V(u)} Q^{u,v} - I)U^\dagger D\|_F^2 = O(\eps)$. We can therefore slightly modify each $Q$
into matrices $Q'$ that sum exactly to the identity on $v$ (recall that $P^{u,v}=0$ whenever $v\notin
N_V(u)$). Now consider the first prover's unitary $\textbf{A}$. Change the basis of $\textbf{A}$ using the
projector's simultaneous diagonalization unitary $U$. Let $\textbf{A'}$ be the matrix with blocs $Q'^{u,v}$.
Fix $v$ and consider the set of lines of $\textbf{A}$ corresponding to this $v$. Since $\textbf{A}$ is
unitary, each of these lines has norm $1$. Moreover by (\ref{eq:diagonal}) they are close to the
corresponding lines of $\textbf{A'}$, which have coefficients in $\{0,1\}$. Therefore these lines can be
slightly modified to have exactly one $1$ per line, yielding matrices $Q''^{u,v}$ that sum to the identity on
$u$, and are still close to the original $Q^{u,v}$.

\end{proof}

\section{Conclusion and future work}\label{sec:rest}

We have attempted to devise a test (our \BIJ) which forces the provers to implement a bijection on the
message register. Obviously the bottleneck to decreasing further the soundness of our protocol is the
increase in error when we go from almost commuting matrices to almost diagonal matrices. The question of how
well almost commuting matrices can be approximated by diagonal matrices has been studied extensively in the
theory of operator algebras, albeit mostly when the norm in question is the operator norm, and not the
Frobenius norm. One might be tempted to conjecture that sets of almost commuting self-adjoint matrices can be
perturbed slightly to a commuting set (that they ``nearly" commute). In fact for the case of just two
matrices, this was a famous conjecture by Halmos \cite{Halmos:conjecture} ({\em Are almost commuting
Hermitian matrices nearly commuting?}). It is known that this conjecture is wrong for two {\em unitary}
matrices: Voiculescu \cite{Voiculescu:ex} gave an example of two unitary $n$-dimensional matrices $A$ and $B$
such that $\|AB-BA\|_{op} \leq 1/n$ but for all commuting $A',B'$ we have $\|A-A'\|_{op}+\|B-B'\|_{op} \geq
1-1/n$. The proof of the latter inequality depends on the second cohomology of the two-torus.
Halmos' conjecture was disproved in the case of three self-adjoint matrices. Finally Halmos' conjecture was
proved by Lin \cite{Lin:commuting} by a "long tortuous argument" \cite{Szarek:survey} using von Neuman
algebras, almost $20$ years after the conjecture had been publicised.

In the case of projectors the Halmos' conjecture is easy to prove, both in the operator and in the Frobenius
norm. This is due to the fact that any two projectors have a common basis in which they are block-diagonal
with at most $2$-by-$2$ blocks. It is tempting to conjecture that Lemma \ref{lem:diagonal} holds with
constant increase in the error. We give here an example, due to Oded Regev, that gives evidence that
Conjecture \ref{conj} might be false for $\delta =O(\sqrt{m})\eps$.

\paragraph{Candidate counterexample:} Let $D$ be always a multiple of $I$ such that $\|D\|_F=1$ ($D$'s dimensions
will adapt to the dimensions of the matrix it is beeing multiplied by) and
\be
 I=\left(\begin{array}{cc} 1 & 0\\0 & 1\end{array}\right) \quad \quad
  Z=\left(\begin{array}{cc} 1 & 0\\0 & -1\end{array}\right) \quad \quad
  W=\left(\begin{array}{cc} 1-\eps & \eta\\ \eta & \eps-1\end{array}\right) \nonumber
\ee where $\eta= \sqrt{(2-\eps)\eps}$, such that $W$ has eigenvalues $1$ and $-1$. As eigenvalues multiply
when matrices are tensored, we have that any tensor product of $n$ of these matrices (of dimension $N=2^{n}$)
has exactly half eigenvalues $1$ and half $-1$. To any such tensor product we will add $I^{\otimes n}$ and
divide by $2$ to make it a projector of rank $2^{n-1}=N/2$. Note that the commutator of two such projectors
equals the commutator of the two tensor products. We omit the $\otimes$ and write e.g. $IIIZW$ for $I \otimes
I \otimes I \otimes Z \otimes W$. We call the first tensor factor {\em position} $1$, the second {\em
position} $2$ and so on, so $IIIZW$ has a $Z$ in position $4$. The {\em weight} of such a tensor product is
the number of positions different from $I$; so the weight of $IIIZW$ is $2$.

We  construct a set of $m$ such tensor products  of weight $\sqrt{m}$ with the property that any two of them
{\em intersect} only in at most one  position, where {\em intersect} in position $i$ means that both matrices
have a tensor factor different from $I$ in position $i$. Note that the norm of the commutator of any two
tensor products that intersect in one position is equal to the norm of the commutator of the matrices in this
position. For example $\|[ IWZZ,IWIW]D\|_F=\|(IWZ \otimes [Z,W])D\|_F=\|[Z,W]D\|_F$. We have $\|[Z,W]D\|_F^2
\leq 8 \eps$.

Choose $m$ such that $\sqrt{m}$ is a prime. Let us arrange the $m$ positions in a square of length
$\sqrt{m}$. Each projector has $I$ everywhere except on a line (modulo $\sqrt{m}$), where its weight is concentrated.
Note that every two lines intersect in at most $1$ position and that there are at least $m$ such lines
($\sqrt{m}$ for each of the $\sqrt{m}$ ``angles"). For the positions on the line let us randomly pick $Z$ and
$W$ with probability $1/2$ each.

We would like to show that there is a {\em good} basis, i.e. a basis in which all the projectors are roughly
diagonal. Given a projector $P$ with, say, a $Z$ in position $i$, there are several other projectors that
intersect with $P$ in $i$ and about half of them will have a $W$ in position $i$. So the good basis that we
are looking for must lie somewhere ``between" $Z$ and $W$. But since this is true for all the positions where
$P$ is different from $I$, there are about $\sqrt{m}/2$ matrices that are {\em misaligned} with $P$. No
matter what basis we finally chose, as long as it is a tensor-product basis, $O(\sqrt{m})$ of the positions
will have something of the form $\pm (1-\eps/2)$ (roughly) on the diagonal. This means that the weight on the
diagonal is roughly $ (1-\eps/2)^{\sqrt{m}} \approx 1-\sqrt{m} \eps$ and hence the off-diagonal weight is
$O(\sqrt{m \eps})$ and hence $\delta=\Omega(\sqrt{m}\eps)$. This is true when the good basis has a tensor
structure, at least, but our search for other good bases has not been successful.

\medskip

Two avenues remain: it might be that the projectors that arise in our proof system have a special structure
which allows to prove approximate diagonalization without too much increase in error. Or else it could be
that Conjecture \ref{conj} is true for some $\delta=poly(n) \eps$, or even constant $\delta$. In the latter
case this would mean that there is some good non-tensored basis for our counterexample.

We have proved our results for a ``scaled down" version, where the verifier has logarithmic workspace and the
quantum messages exchanged have a logarithmic number of qubits. It is possible to scale up these results: by
carefully choosing a \NEXP-complete version of \GM, with $|U|=|V|=|W|=2^n$ and $|M|=O(2^n)$, such that the
degree remains constant, our proof works with messages of length $O(n)$ and a polynomially bounded verifier
to imply $\NEXP \subseteq \QMS_{1,s}(2,1)$ with soundness $s$ doubly exponential in $n$. Note that in this
case the verifier cannot read his input in polynomial time. However, given $u\in U$ he only needs to be able
to find all (constantly many) $(v,w)\in V\times W$ such that $(u,v,w)\in M$. The details of this construction
will be given in an ulterior version of this paper.

We hope that our proof technique will be useful in other contexts. For instance one could imagine using it
to give quantum interactive protocols for other problems, both \NP-complete or not. Preliminary attempts have
shown that similar techniques work to give \QMS-protocols for {\sc 3COLORING}. Or one could try to give
quantum interactive protocols for problems that are between \Pe\ and \NP-complete, and base $\QMS \nsubseteq
\EXP$ on the hardness of those.

\section{Acknowledgments}
We thank Oded Regev and Ben Toner for extended discussions on $\QMS$ and $\MIP^*$ and for generously sharing
their knowledge with us, and Oded for providing the candidate counterexample. We thank Umesh Vazirani for
very useful discussions during earlier work involving one quantum prover. We also thank Stanislav Szarek for
discussions about almost commuting and almost diagonal matrices.

\newcommand{\etalchar}[1]{$^{#1}$}


\end{document}